\documentclass[12pt]{iopart}
\usepackage{graphicx}

\begin{document}

\title[Sub-Poissonian  and  anti-bunching via majorization]{Sub-Poissonian and anti-bunching criteria via majorization of statistics}

\author{Daniel G. Medranda and Alfredo Luis}

\address{Departamento de \'{O}ptica, Facultad de Ciencias
F\'{\i}sicas, Universidad Complutense, 28040 Madrid, Spain}
\ead{alluis@ucm.es}
\vspace{10pt}
\begin{indented}
\item[] \today
\end{indented}

\begin{abstract}
We use majorization and confidence intervals as a convenient tool to compare the 
uncertainty in photon number for different quantum light states. To this end majorization 
is formulated in terms of confidence intervals.  As a suitable case study we apply this 
tool to sub- and super-Posissonian behavior and bunching and anti-bunching effects.  
We focus on the most significant classical and nonclassical states, such as Glauber 
coherent, thermal, photon number, and squeezed states. We show that majorization 
provides a more complete analysis that in some relevant situations contradicts the 
predictions of variance.
\end{abstract}

%
\vspace{2pc}

\section{Introduction}

The most typical measure of uncertainty is variance, both in classical and quantum physics. 
However, variance is not the only  measure of fluctuations  nor the best behaved, as already 
pointed out for example in Refs. \cite{AV}.  Some well-known disadvantages are, especially 
from a quantum perspective:
(i) May fail when we move away from Gaussian-like statistics. In contrast to classical  physics, 
non-Gaussian statistics is often the case in quantum physics. 
(ii) Puts too much emphasis on extremely unlikely outcomes. This may be a serious drawback 
in practical applications where the knowledge of the probability distribution is not complete.
(iii) It cannot be used to derive Heisenberg-like uncertainty relations in finite-dimensional 
systems.

This has motivated the introduction of several measures of quantum fluctuations alternative to 
variance \cite{MO,AD,RSS,AR,CT,CB,UH}. Most of them have the form of an entropy, such as the 
R\'{e}nyi and Tsallis entropies, that include the Shannon entropy as a particular case. 
At the very heart of these alternative entropic measures lies the concept of majorization
since it is equivalent at once to the most popular entropic measures that satisfy the property 
of {\em Schur-concavity}. 

In this work we address the majorization of photon-number statistics in quantum optics. 
We put a particular emphasis on the comparison with Poissonian statistics. The reason for this 
is due to the fact that the comparison of variances is the most classic test disclosing nonclassical 
behavior in quantum optics and plays a relevant role in quantum metrology \cite{SZ}. The most 
significant advances presented  in this contribution are: 

(i) We show that majorization is equivalent to the unanimity of confidence intervals.
Confidence intervals. They have been already used  as a suitable measure of quantum 
uncertainty \cite{UH}. This extends the equivalence of majorization to other measures 
of uncertainty beyond entropies.

(ii) We show that majorization provides a much more stringent definition of sub- and 
super-Poissonian statistics, that may not coincide with the variance-based definition. 

(iii) We extend the analysis to cover bunching and anti-bunching effects. Using variance 
as a measure of comparison they are equivalent to super- and sub-Poissonian statistics, respectively. 
We introduce a suitable criterion for bunching and anti-bunching in terms of majorization  
as a more complete statistical assessment. We show that from this perspective the 
equivalence with super- and sub-Poissonian statistics no longer holds and may provide 
a new practical criterion to disclose nonclassical behavior.

(iv)  We apply this approach to the most frequent classical states, such as coherent and 
thermal states, as well as the most common nonclassical states,  such as  number states 
and squeezed states. All these states can be generated experimentally  and have many 
practical applications.

The deeper statistical characterization provided by  majorization can have interesting 
physical consequences in quantum physics and quantum optics in particular since
fluctuations and uncertainty are a cornerstone of the quantum theory. For the 
sake of illustration let us focus on possible physical consequences in the area of quantum 
metrology.

One of the objectives of quantum metrology is to determine whether quantum fluctuations 
limit the precision in the detection of weak signals. Currently, most researchers believe 
that such limits exist under the form of Heisenberg  limit \cite{HL}. However, such belief 
is based on variance estimators, so alternative approaches to quantum uncertainty may 
eventually be useful to overcome their metrological effects, as already suggested in 
Ref.  \cite{LR}. The probe states with sub-Poissonian photon statistics are crucial to reach 
the ultimate quantum resolution. Therefore, a more complete understanding of sub-Poissonian 
statistics may shed light on these limits and their eventual breaking. 

Sub-Poissonian photon statistics is a paradigmatic nonclassical effect revealed by the comparison 
between variances. In this context, majorization is a comparison between complete distributions, 
so it may provide new tests of nonclassical behavior with a deep relation with previous  nonclassical 
criteria introduced in terms of photon-number moments higher than variance \cite{HON,LEE}.  
Actually, majorization has been already used to derive higher-order criteria of nonclasssical behavior 
in Ref. \cite{LEE}.

Deep down, we can regard quantum-metrology limits as the consequence of the uncertainty 
relations satisfied by complementarity observables, such as time and energy. Uncertainty 
relations are typically expressed in terms of variance, so it may be interesting to investigate 
the consequences that might be derived from uncertainty relations expressed in terms of 
generalized entropies, or even via majorization such as in Ref. \cite{NQUR}.  For example, 
there is no position-momentum uncertainty relation if uncertainty is assessed using some 
particular entropic measures \cite{ZPV}. Moreover, it has been shown that two observables 
can be complementary or not depending on the measure of fluctuations used \cite{AL02}, 
and nontrivial uncertainty  measures can solve long standing debates on the enforcement of 
complementarity \cite{LS}. 

Finally, quantum metrology involves taking decisions with incomplete information. In such 
conditions it is preferable to work directly with probabilities rather than with their moments 
in order to obtain meaningful estimators of uncertainty.

In Sec. 2 we recall the concept of majorization, relating it with confidence intervals as suitable 
uncertainty measures, as well as  most common photon statistics. We compare different photon 
statistics in Sec. 3.  In Sec. 4 we extend these ideas to bunching and anti-bunching phenomena. 

\section{Majorization, uncertainty and photon statistics}

In this section we just recall the concept of majorization, its relation with uncertainty measures, 
and the photon statistics of the most usual and interesting classical and nonclassical states.

\subsection{Majorization}

The distribution $p_a$ majorizes  the distribution $p_b$, which will be expressed as $p_b \prec 
p_a$, when the following relation between the ordered partial sums $S_N$ holds,  as illustrated 
in Fig. 1
 \begin{equation}
 \label{major}
S_N  \left ( p_b^\downarrow \right ) =  \sum_{j=0}^N p^{\downarrow} _{b,j} \leq \sum_{k=0}^N
p^{\downarrow} _{a,k} = S_N  \left ( p_a^\downarrow \right  )  , 
 \end{equation}
 for $N=0,1 \ldots \infty$, where the superscript $\downarrow$ on $p^\downarrow$ denotes the 
 same distribution $p$ after a permutation of $p_n$  rearranging them in decreasing 
 order 
\begin{equation}
p^{\downarrow} _1 \geq p^{\downarrow} _2 \geq \ldots \geq p^{\downarrow} _j \geq \ldots .
\end{equation}
Majorization is a partial-ordering relation, so there are distributions that neither $p_b \prec p_a$ 
nor $p_a \prec p_b$. 

\begin{figure}
\begin{center}
\includegraphics[width=6cm]{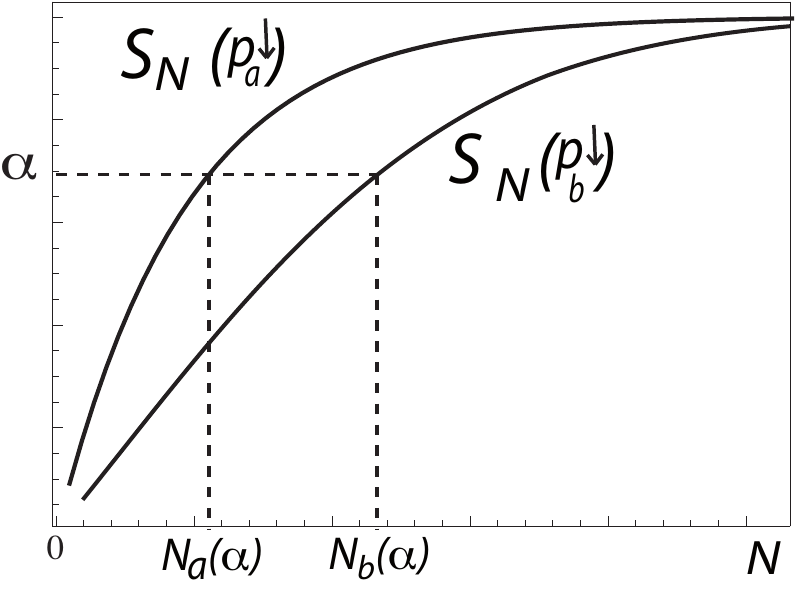}
\caption{Relation between partial ordered sums $S_N$ and confidence intervals 
when the majorization  $p_b \prec p_a$ holds.}
\end{center}
\end{figure}

\subsection{Uncertainty, entropies and confidence intervals}

Roughly speaking, if  $p_a$ majorizes $p_b$ we have that $p_b$  is more disordered or more 
mixed  than $p_a$. In other words, $p_a^\downarrow$ is more concentrated around its maximum. 
This can be made more precise by noting that majorization is equivalent to a relation between 
statistics in terms of doubly stochastic matrices \cite{MO}. This means that $p_b$ can be obtained 
from $p_a$ by the addition of classical noise.

Thus majorization provides a very convenient measure of uncertainty. For example, the distribution 
with full certainty  $p^{\downarrow} _1 =1$, $p^{\downarrow} _{j \neq 1} =0$ has $S_N = 1$ for all 
$N$,  and majorizes all distributions. In the case of photo number, the states with full certainty 
are the photon-number states $| n\rangle$.  

Whenever two distributions can be compared via majorization, the result is respected by the 
Schur-concave entropies, such as Tsallis and R\'{e}nyi entropies:
\begin{equation}
R_q (p_i) = \frac{1}{1-q} \ln \left ( \sum_i p_i^q\right ), \qquad
T_q (p_i) = \frac{1}{1-q} \left ( \sum_i p_i^q - 1\right ),
\end{equation}
where $q \geq 0$ is the entropic index. The limiting case $\alpha \rightarrow 1$  is the Shannon 
entropy $T_1=R_1 =  - \sum_i p_i \log p_i$. Majorization implies that if  $p_b \prec p_a$ then 
$H_q (p_b ) > H_q (p_a)$ for all $q$ and $H = T,R$.

As an alternative and more direct connection of majorization with uncertainty measures, let us 
consider the confidence intervals $N (\alpha)$. They are defined as the minimum number of  
indices such that  the partial sum up to $p^\downarrow_N$ comprises a given fraction $\alpha$ 
of the probability 
\begin{equation}
 \sum_{j=0}^N p^{\downarrow} _j \geq \alpha , \quad N \geq N (\alpha) .
 \end{equation}
When two distributions are comparable, $p_b \prec p_a$ is equivalent to saying that all confidence 
intervals of $p^\downarrow_b$ are larger  than or equal to  those of $p^\downarrow_a$:
\begin{equation}
\label{cond}
p_b \prec p_a \longleftrightarrow N_b (\alpha) \geq N_a (\alpha) \; \;  \forall \alpha .
\end{equation}
Otherwise, if the distributions are incomparable we will have $ N_b (\alpha) >  N_a (\alpha)$ and  
$N_b (\beta) < N_a (\beta)$ for some $\alpha$, $\beta$.

\subsection{Photon number statistics}

Let us recall  the photon-number statistics of  the most significant classical and nonclassical 
states beyond the trivial case of photon-number states.

\subsubsection{Coherent light}

The most celebrated light states regarded as classical in quantum optics are the Glauber coherent 
states. They have a Poissonian photon number distribution $p_{c,n}$
\begin{equation}
p_{c,n} = \frac{\bar{n}^n}{n!} \exp \left (-\bar{n} \right ) , \qquad \Delta^2 n = \bar{n} .
\end{equation}
This depends on a single parameter, the mean number of photons $\bar{n}$, that equals the 
variance $\Delta^2 n$. Unfortunately, there is no suitable expression for  $p_{c,n}^\downarrow$, 
nor for the partial sums $S_N (p_{c,n}^\downarrow)$. Nevertheless, in the case $\bar{n} \gg 1$ 
the Poissonian distribution can be well approximated by a Gaussian, treating $n$ as a continuous 
variable
\begin{equation}
p_{c,n} \simeq p_{G,n} = \frac{1}{\Delta n \sqrt{2 \pi}} \exp \left [ - (n-\bar{n} )^2/ (2 \Delta^2 n ) \right ] .
\end{equation}
In this case $p_{G,n}^\downarrow$ can be easily obtained because of the full symmetry and 
monotonic behavior around the maximum at  $\bar{n}$. The corresponding partial ordered sums are
\begin{equation}
\label{SNG}
S_N  \left ( p_{G,n}^\downarrow \right ) =  \sum_{n=0}^N p^{\downarrow} _{G,n} = \frac{2}{\sqrt{\pi}} 
\int_0^\frac{N}{\sqrt{2} \Delta n}
\rmd u \exp \left ( -u^2 \right ).
\end{equation}

\subsubsection{Thermal light}

Another basic and very common example of classical light are the thermal states, with photon-number 
statistics
\begin{equation}
p_{t,n} =  \frac{1}{\bar{n} + 1} \left ( \frac{\bar{n}}{\bar{n} + 1} \right )^n , \quad \Delta^2 n = \bar{n} 
( \bar{n} + 1),
\end{equation} 
that again depends on a single parameter. All thermal states are super-Poissonian $\Delta^2 n > 
\bar{n}$. The probability distribution is in decreasing order from the start $p_{t,n}^\downarrow = 
p_{t,n}$, and the partial ordered sums can be easily obtained analytically as 
\begin{equation}
\label{SNt}
S_N  \left ( p_{t,n}^\downarrow \right ) =  \sum_{n=0}^N p^{\downarrow} _{t,n} =  1 - \left ( 
\frac{\bar{n}}{\bar{n} + 1} \right )^{N+1} .
\end{equation}
 
\subsubsection{Squeezed light}

This is may be the best-known practical example of nonclassical light beyond photon-number states. 
Squeezed states are obtained from the vacuum $|0 \rangle$ by a squeezing transformation $S(r)$ 
followed by a coherent displacement $D(R)$ as $D(R)S(r) |0 \rangle$.  For simplicity we will consider 
squeezing parameters such that the photon-number distribution is  \cite{SZ}
\begin{eqnarray}
& p_{s,n} = \frac{\tanh^n r}{2^n n! \cosh r} \exp \left [ - R^2  \left ( 1 - \tanh r \right ) \right ] & \nonumber \\
& \times \left | H_n \left [ \frac{\rmi R}{\sqrt{2}} \left (  \sqrt{\tanh r} - \frac{1}{\sqrt{\tanh r}} \right ) \right ] \right |^2 , &
\end{eqnarray}
where $R$ represents the coherent contribution while the squeezing parameter $r$, with sign, is the 
squeezed part that represents the squeezing of quadrature fluctuations. The mean value and 
variance of the photon number are
\begin{equation}
\label{mav}
\bar{n} = R^2 + \sinh^2 r, \qquad \Delta^2 n = R^2 \rme^{2r} + \frac{1}{2} \sinh^2 (2r) .
\end{equation} 
They present sub- or super-Poissonian statistics depending on the $R,r$ values. In any case there is 
no simple form neither  for $p_{s,n}^\downarrow$ nor for the ordered partial sums 
$S_N (p^\downarrow_{s,n})$.

\subsection{Majorization applied to sub- and super-Posissonian statistics}

The idea of sub- and super-Posissonian statistics results from the comparison of the photon-number 
distribution $p_n$ of a given field state with a  Poissonian statistics $p_{c,n}$ of the same mean 
number. This comparison has been always done in terms of variance. Majorization provides a much 
more stringent criterion as follows:
\begin{equation}
\label{Pc}
p \prec p_c \equiv \mathrm{under \mbox{-} Poissonian}, \qquad p_c \prec p \equiv 
\mathrm{over \mbox{-} Poissonian} ,
 \end{equation}
where in order to avoid confusions we have provided new different names. Over-Poissonian means 
that the curve $S_N (p^\downarrow )$ is always above the curve $S_N (p_c^\downarrow )$ and vice 
versa for under-Poissonian. For Gaussian statistics over-Poissonian is equivalent to the sub-Poissonian 
and under-Poissonian equivalent to super-Poissonian. Otherwise, if the distributions are incomparable
there can be no unambiguous claim about their relative photon-number uncertainty, and the result 
will depend on the particular entropy  $H_q$ or confidence value $\alpha$ employed. 

\section{Majorization relations}

Next we  compare via majorization the photon-number distributions for the above classical and nonclassical 
states. When contrasting states from different families we will tend to compare states with the same mean 
number of photons. The reason to proceed in this way is that we are focusing on sub-Poissonian behavior, 
which implies comparison with a Poissonian distribution of the same mean.

\subsection{Coherent versus coherent}

 It has been shown in Ref. \cite{MO} that any two Poisson distributions can be related through a doubly 
stochastic matrix in such a way that the distribution with smaller mean (this is smaller variance) majorizes the 
distributions with larger mean.  Therefore, the order established by majorization coincides with the order 
established by variance and the distribution with smaller $\Delta n$ always majorizes the distribution with larger 
$\Delta n$. As illustration in Fig. 2 we have represented the ordered partial sums $S_N  \left ( p_{c,n}^\downarrow 
\right ) $ as functions of $N$ for coherent states with  different mean number of photons. This result is especially 
meaningful for small mean numbers. Otherwise,  for  $\bar{n} \gg1 $  the Gaussian approximation holds and 
majorization readily follows from variance after  Eq.  (\ref{SNG}). 
 
\begin{figure}
\begin{center}
\includegraphics[width=8cm]{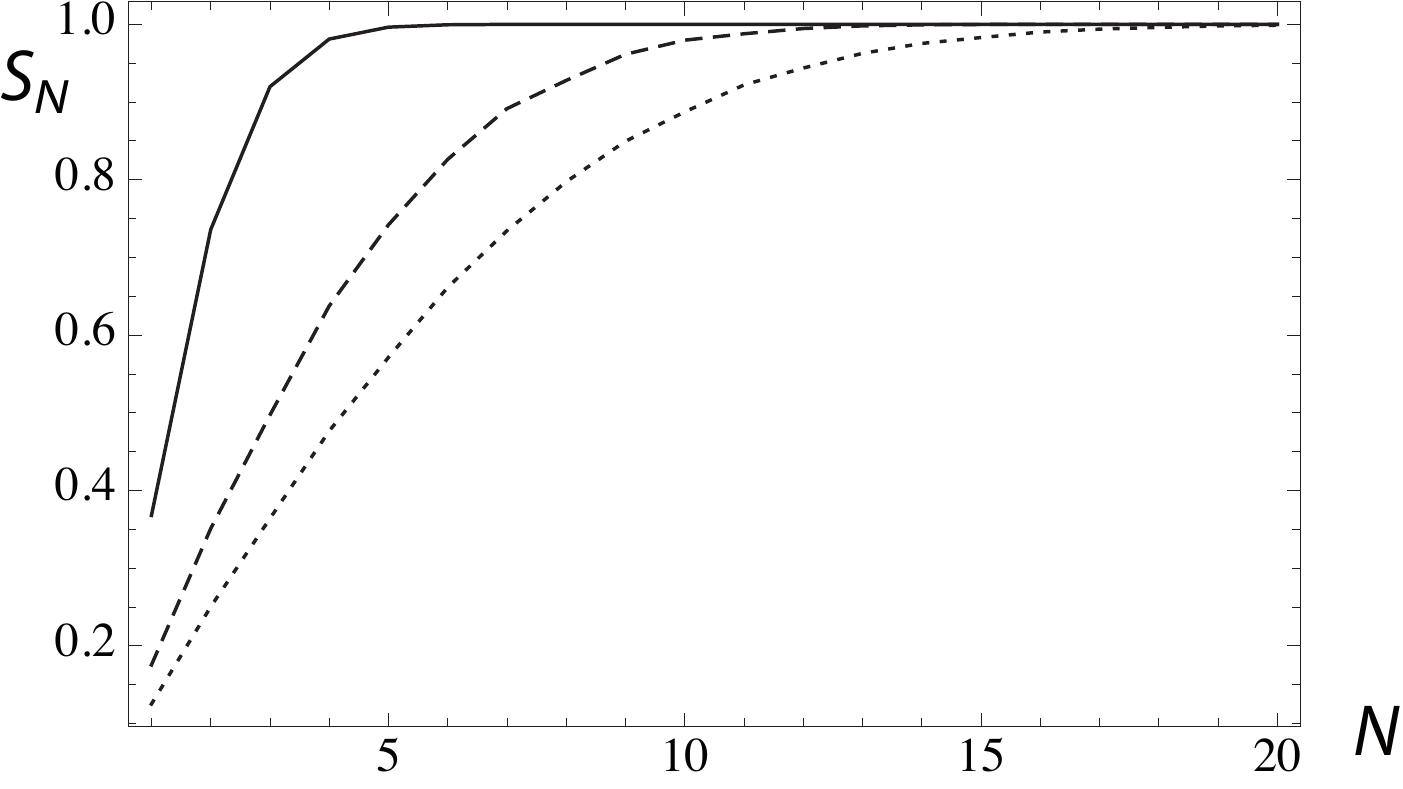}
\caption{Ordered partial sums $S_N  \left ( p_{c,n}^\downarrow \right )$ as functions
of $N$ for coherent states with mean number of photons (from top to bottom) $\bar{n} =1,5,10$. 
Throughout lines are drawn as guides for the eye.}
\end{center}
\end{figure}

\subsection{Thermal versus thermal}

An equivalent majorization relation consistent with variance follows for thermal light directly from 
relation (\ref{SNt})
\begin{equation}
S_N  \left ( p_{t,a}^\downarrow \right ) > S_N  \left ( p_{t,b}^\downarrow \right ) 
\longleftrightarrow \Delta_{t,a} n < \Delta_{t,b} n .
\end{equation}
This case has been  examined in more detail in Ref. \cite{MP}. In this case majorization may 
be pictured also as the result of  added noise after Eq. (15)  in Ref. \cite{HA} for example.

\subsection{Thermal versus coherent}
Using variance as a measure of comparison, the thermal light should always have larger uncertainty 
than coherent light for the same mean number.  However we have found that they can be incomparable 
for small mean photon numbers, such as for $\bar{n}_t = \bar{n}_c = 1.5$, as shown in Fig. 3.  
The ordered partial sums cross at a critical value $\tilde{\alpha} \simeq 0.80$ for $N ( \tilde{\alpha} ) 
\simeq 3$. Confidence intervals $N (\alpha < \tilde{\alpha} )$ say that  thermal light has less number 
fluctuations than coherent light, i. e., behaves as over-Poissonian light, while the confidence 
intervals $N (\alpha > \tilde{\alpha} )$ say the opposite and the thermal behaves as under-Poissonian. 
The transition occurs for a confidence interval that for Gaussian statistics would be  between 
one and two standard deviations.  On the other hand, for  all cases examined with $\bar{n}_c =  
\bar{n}_t  \gg 1$ the coherent states always majorize the thermal light, in accordance with variance. 

When we remove the condition of equal mean, we can  get  incomparable distributions even in 
the case of large mean numbers $\bar{n}_c , \bar{n}_t  \gg 1$. Moreover, in such a case  we can find 
examples of full contradiction between majorization and variance, for  example for  $\bar{n}_c = 100$ 
and $\bar{n}_t = 10$.  The variance of the thermal state is larger than the  coherent, $\Delta^2_t n = 110$ 
versus $\Delta^2_c n = 100$, while the thermal almost majorizes the coherent, as shown in Fig. 4. Strictly 
speaking, the ordered partial sums cross at $\tilde{\alpha} \simeq 0.995$ for $N (\tilde{\alpha}) \simeq 57$ 
as shown in the inset of Fig. 4. However, the 99.5 \% of the photon-number statistics, and thus almost all 
confidence intervals $N (\alpha )$, support that the thermal state presents lesser uncertainty 
than coherent light, contradicting variance. This result clearly illustrates how variance overestimates 
outcomes with extremely small probability, such as $0.5 \%$ in our case.

\begin{figure}
\begin{center}
\includegraphics[width=8cm]{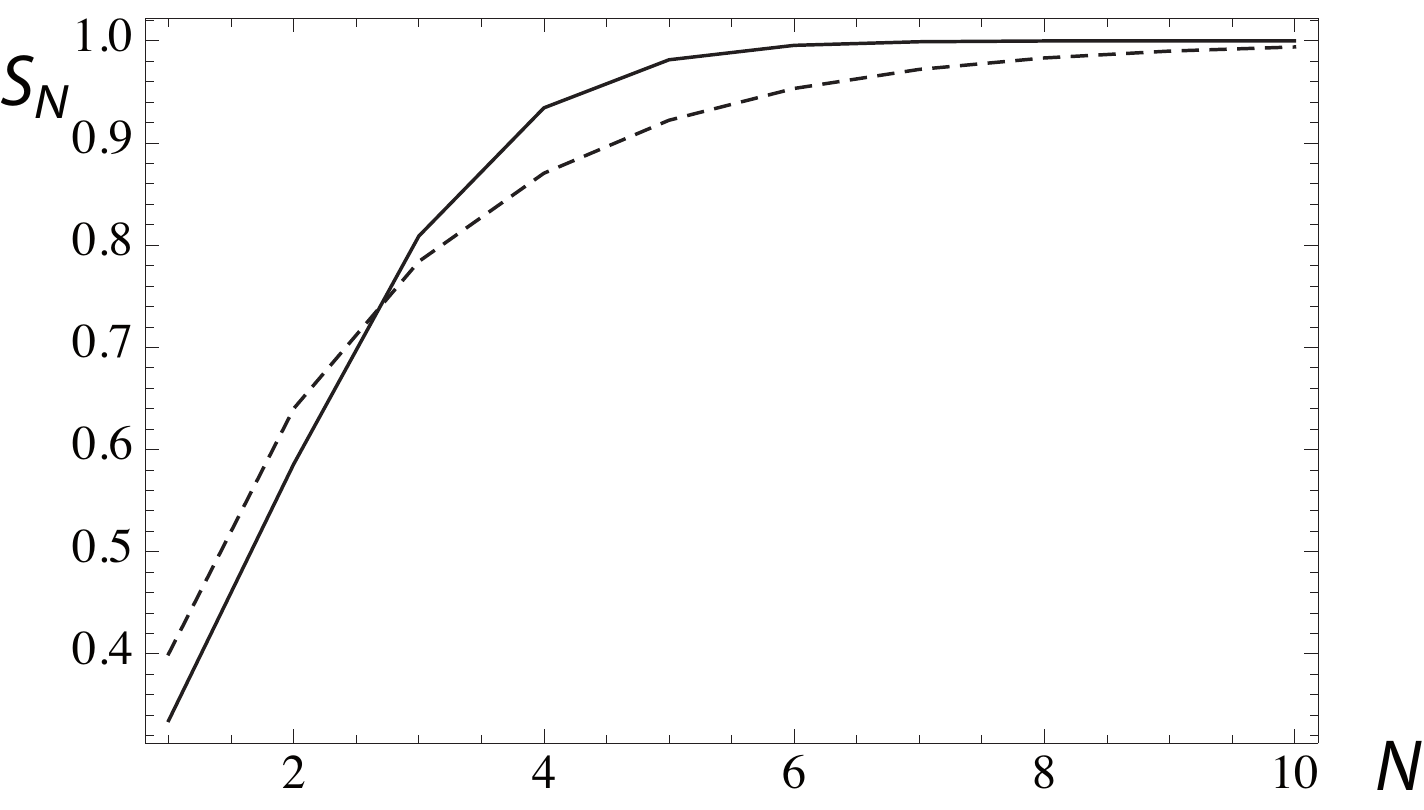}
\caption{Ordered partial  sums $S_N  \left ( p_{c,n}^\downarrow \right ) $ and  $S_N  \left ( 
p_{t,n}^\downarrow \right ) $  as functions of $N$ for coherent  (solid) and thermal (dashed)  light 
with mean number of photons $\bar{n}_t = \bar{n}_c = 1.5$. }
\end{center}
\end{figure}

\begin{figure}
\begin{center}
\includegraphics[width=8cm]{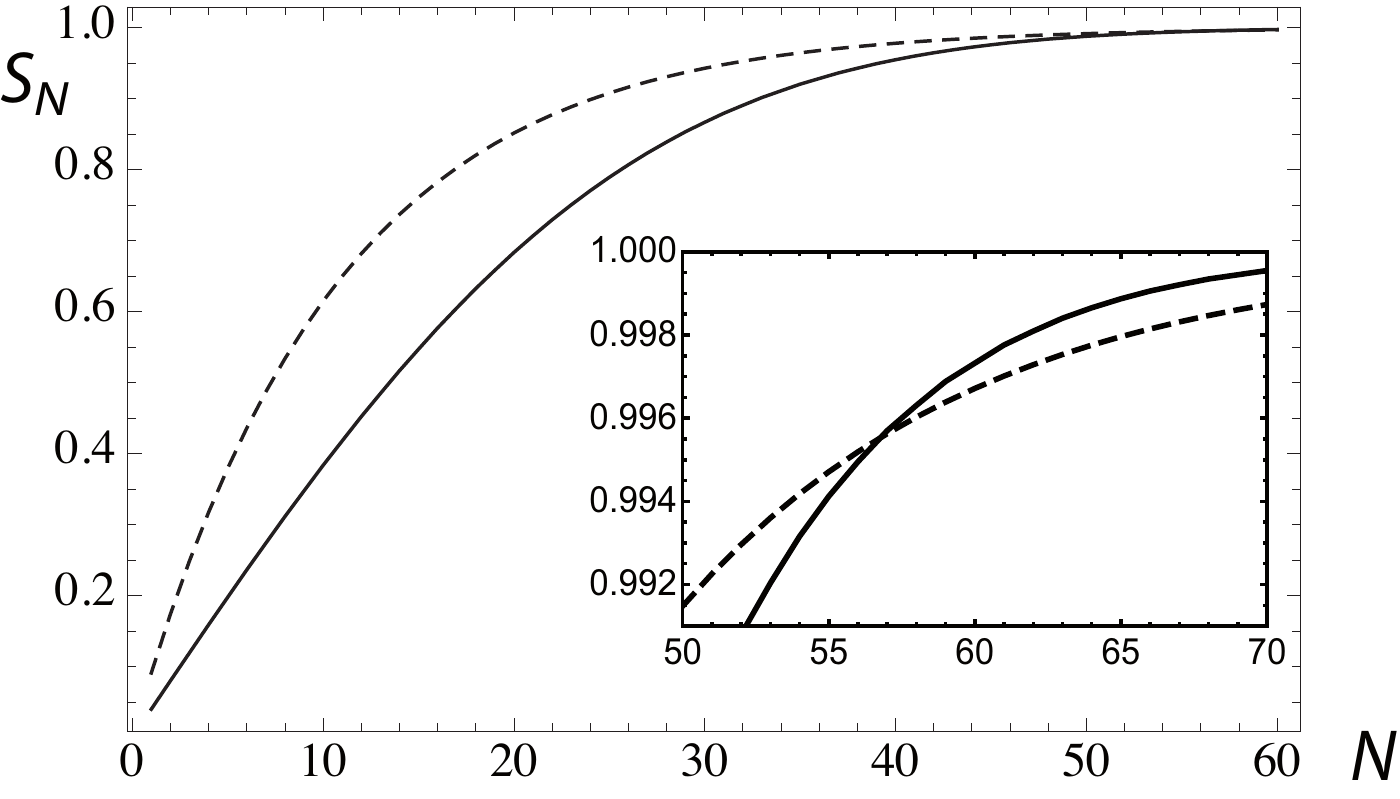}
\caption{Ordered partial sums  $S_N  \left ( p_{c,n}^\downarrow \right ) $ and  $S_N  \left ( 
p_{t,n}^\downarrow \right ) $ as functions of $N$ for coherent (solid) and thermal (dashed)  light 
with mean number of photons $\bar{n}_t = 10$ and $\bar{n}_c = 100$. The inset shows an 
extension and detail  of the plot for larger $N$ revealing  the crossing of the curves. }
\end{center}
\end{figure}

\subsection{Squeezed versus squeezed and coherent}

\begin{figure}
\begin{center}
\includegraphics[width=8cm]{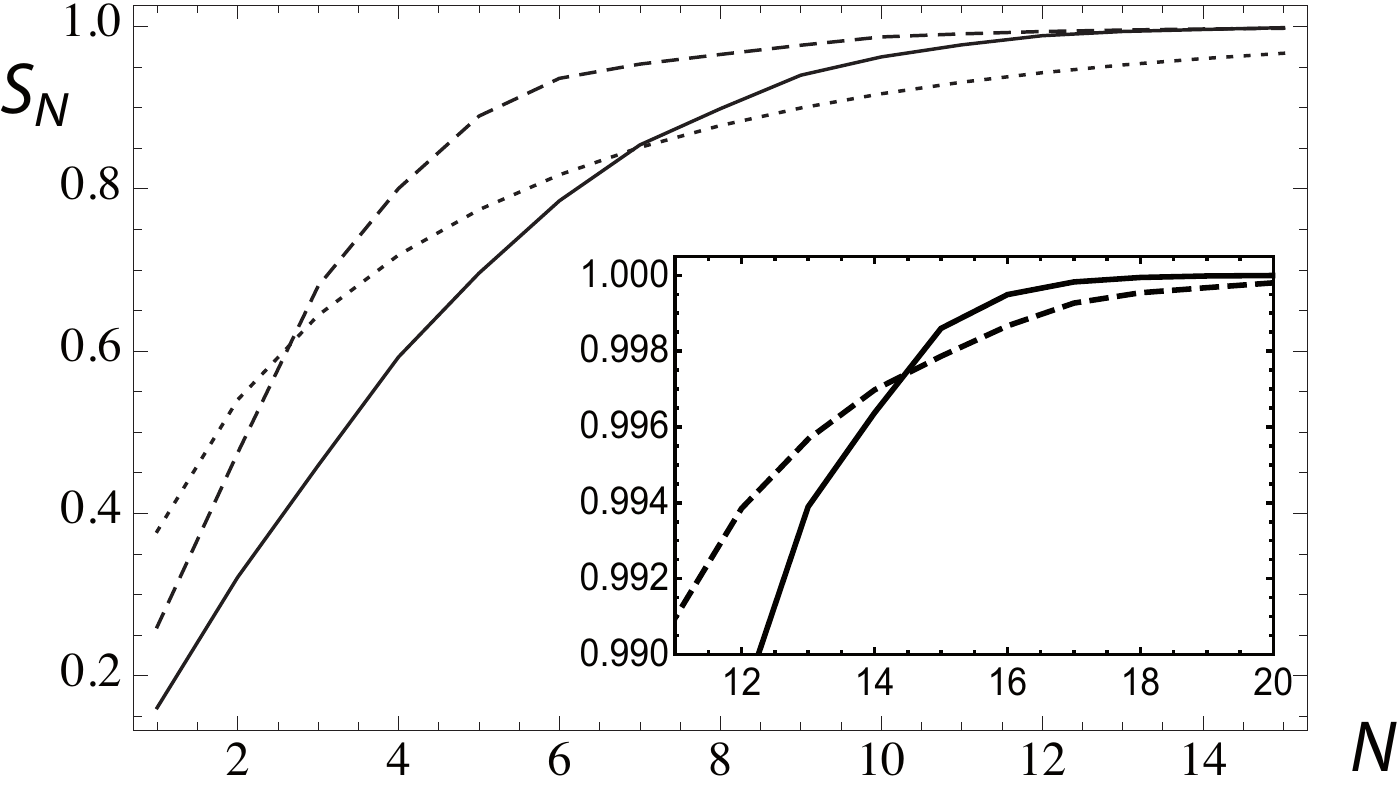}
\caption{Ordered partial sums for variance-like super-Poissonian squeezed state with 
$\Delta_\mathrm{super}^2 n = 14 \, \bar{n}$ (dotted), sub-Poissonian squeezed state  
with $\Delta_\mathrm{sub}^2 n = 0.6 \, \bar{n}$ (dashed) and coherent light 
$\Delta_c^2 n =\bar{n}$ (solid) as functions of $N$ with $\bar{n}=6$. The inset shows 
an extension and detail  of the plot for larger $N$ values revealing the crossing of the 
sub-Poissonian and coherent curves. }
\end{center}
\end{figure}

In the case of squeezed states we will consider always the same mean number of photons 
$\bar{n}$ for all states, but different number variance for some particular combinations of 
$R$ and $r$ Eq. (\ref{mav}). We have found incomparable statistics when contrasting 
sub-Poissonian states, when comparing  super-Poissonian states, as well as when 
comparing them with Poissonian light.

An example of incomparable states is illustrated in Fig. 5 where  we have plotted the ordered 
partial sums $S_N ( p_n^\downarrow )$ for three different light states with $\bar{n}=6$ and 
variances $\Delta_\mathrm{sub}^2 n = 0.6 \, \bar{n}$  (dashed),  $\Delta_\mathrm{super}^2 
n = 14 \, \bar{n}$  (dotted), and  $\Delta_c^2 n = \bar{n}$ (solid). The super-Poissonian case 
is actually the squeezed vacuum $R=0$. We can appreciate that both squeezed states 
are incomparable and the partial sums cross at $\tilde{\alpha} \simeq 0.60$ for  $N ( \tilde{\alpha} ) 
\simeq 2.6$.  Moreover the coherent and super-Poissonian states are incomparable as well and 
the partial sums cross at $\tilde{\alpha} = 0.85$ for  $N ( \tilde{\alpha} ) \simeq 7$. 

On the other hand, it seems in the main plot of Fig. 5 that the sub-Poissonian light (dashed line)
majorizes the coherent light (solid line). Strictly speaking, this is not exactly the case if we 
look closer to large $N$ values, displayed in the inset in Fig. 5, since the partial sums cross at  
$\tilde{\alpha} = 0.997$ for  $N ( \tilde{\alpha} )  \simeq 14$. This might be related to the 
characteristic photon-number oscillations of squeezed light. Nevertheless the fact that  the 
states are incomparable has no practical consequences since it relies on extremely fine details 
of the photon-number distribution that will be hidden by experimental uncertainty. 

\section{Bunching and anti-bunching}

A simple and direct practical observation of the sub- and super-Poissonian statistics is provided 
by the  anti-bunching and bunching phenomena \cite{MW}. This is investigated by mixing the 
field state $\rho$ with vacuum at a 50 \% lossless beam splitter (see Fig. 6).  At the output beams 
two photodetectors register the output photon numbers $n_1$ and $n_2$. In these conditions it 
can be easily seen that 
\begin{equation}
\label{b1}
\Delta n_1 \Delta n_2 = \overline{n_1 n_2} -  \overline{n}_1 \overline{n}_2 = \frac{1}{4} \left ( 
\Delta^2 n -  \overline{n}  \right ) .
\end{equation}
Therefore, the sub- and super-Poissonian character is reflected unambiguously on the sign of  
$\Delta n_1 \Delta n_2 $: super-Poissonian light produces positive correlations  $\Delta n_1 
\Delta n_2  > 0 $, which is known as bunching, while sub-Poissonian light produces negative 
correlations   $\Delta n_1 \Delta n_2 < 0$, known as anti-bunching. 

\begin{figure}
\begin{center}
\includegraphics[width=5cm]{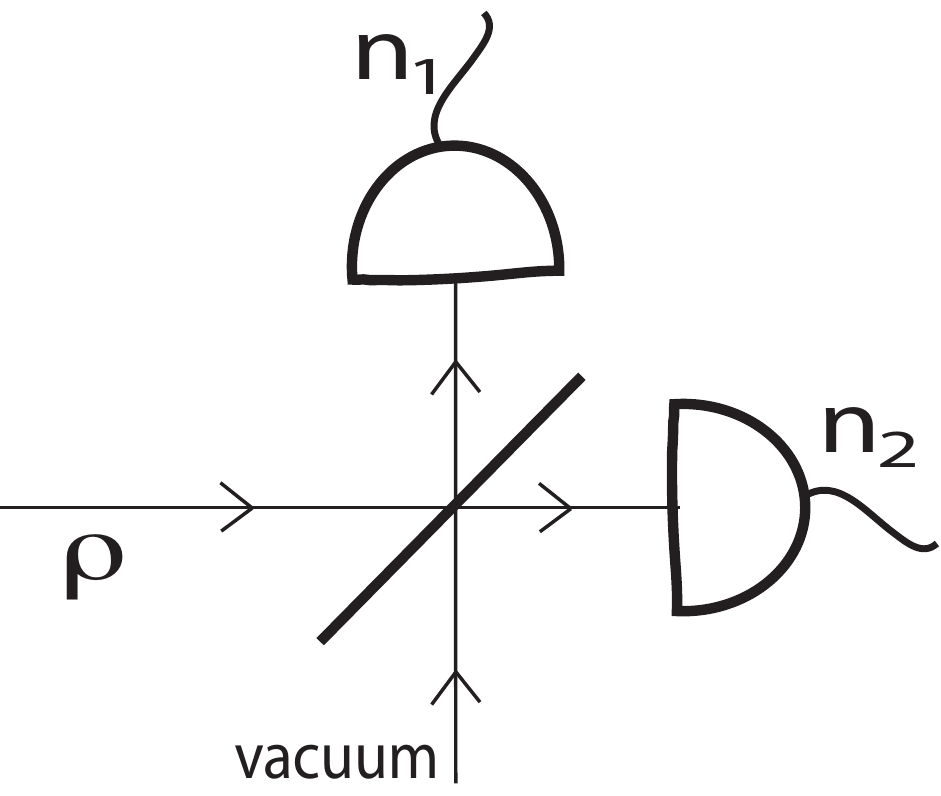}
\caption{Arrangement for the observation of the bunching and anti-bunching effects.}  
\end{center}
\end{figure}

We translate to this framework the  ideas around majorization in order to provide statistical 
characterizations beyond  $\Delta n_1 \Delta n_2 $. This turns out to be quite simple if we focus 
on the number sum $n_+ = n_1 + n_2$ and number difference  $n_- = n_1 - n_2$ since
for these variables we have  
\begin{equation}
\label{b2}
\Delta n_1 \Delta n_2 = \frac{1}{4} \left ( \Delta^2 n_+  -  \Delta^2 n_-  \right ) ,
\end{equation}
so that bunching is equivalent to $\Delta n_+  >  \Delta n_- $ while anti-bunching means $\Delta n_+ 
 <  \Delta n_- $. Because of number conservation and the vacuum input all registered photons 
 $n_1$, $n_2$ come from the state $\rho$ and no photon is lost. Therefore,  the number sum 
 coincides with the number variable $ n_+ = n$ for all $\rho$. Then, after Eqs. (\ref{b1}) and (\ref{b2}) 
 we have  always $\Delta^2 n_- = \overline{n}$, so that the variable $n_+$ carries the number 
 statistics of the state examined while the variable $n_-$ provides a Poissonian reference.

After Eq. (\ref{b2})  we can introduce a more complete criterion beyond variance via a majorization 
comparison between the distributions for the number sum $p_+$ and number difference $p_-$. As
we have just said, the $p_+$  distribution coincides with the number distribution 
of the input estate
\begin{equation}
\label{pp}
p_{+,n} = p_n = \langle n | \rho | n  \rangle  ,
\end{equation}
while for $p_-$ we have 
\begin{equation}
\label{pm}
p_{-,m}= \sum_{n=|m|}^\infty p_n \frac{1}{2^n} \pmatrix{ n \cr \frac{n+m}{2} }  ,
\end{equation}
and we note that when the number sum $n$ is odd or even,  the number difference $m$ is also odd or 
even. Then we can introduce the following criterion
\begin{equation}
\label{bc}
p_+ \prec p_- \equiv \mathrm{clustering}, \qquad p_- \prec p_+ \equiv \mathrm{anti \mbox{-} clustering},
 \end{equation}
where in order to avoid confusions with the variance-based {\em bunching}  we will use the 
name {\em clustering}. Thus clustering means $N_+ ( \alpha ) \geq N_- (\alpha )$ for all $\alpha$ 
and is the majorization counterpart of bunching.  Otherwise, if the distributions are incomparable 
there can be no unambiguous claim, and the result will depend on the particular entropy  $H_q$ or 
confidence value $\alpha$ employed to assess the uncertainty. 

Note that the clustering criterion makes no reference whatsoever to the statistics of coherent 
states, since both  $p_\pm$ in Eqs. (\ref{pp}) and (\ref{pm}) emerge exclusively from the number 
statistics of the state being examined. After Eq. (\ref{b1}) the variance-based bunching and 
anti-bunching are equivalent to super- and sub-Poissonian statistics, respectively. However 
this is not the case of the majorization-based clustering and anti-clustering with regard to 
under- and over-Poissonian statistics. Over-Poissonian means $p_c \prec p $ while anti-clustering 
$p_- \prec  p$. There seems to be no  logical connection between these conditions since $p_-$ 
in general is not Poissonian $p_c$.

Next we look for any relation between anti-clustering and nonclassicality that might 
parallel the well known role played by sub-Poissonian statistics as a nonclassical criterion. There is a 
simple connection if we focus on classical models where vacuum means a wave of zero amplitude. 
In such a case,  after the 50\% beam splitter in Fig. 6 the intensities recorded at both detectors are 
necessarily equal, regardless of the amplitude of the signal wave. This implies that the $p_{-,m}$ 
distribution has all probability concentrated in a single outcome, say $m=0$, and therefore  $p_{-,m}$ 
will majorize any other distribution. Thus, the anti-clustering criterion in Eq. (\ref{bc}) would be a nonclassical 
effect.

As a first and readily example we may consider a coherent state. There is no  simple expression for 
$p_{-,m}$ in Eq. (\ref{pm}) so we resort to a numerical comparison. We have found that for large mean 
numbers, say $\bar{n} \geq 5$, the  partial sums of  $p^\downarrow_{-,m}$ and 
$p^\downarrow_{+,n}$ are almost indistinguishable. On the other hand, for small mean numbers the 
situation is slightly different as shown in Fig. 7 for $\bar{n} = 1$. The distributions turn out to be 
incomparable while all the partial sums of $p^\downarrow_{-,m}$ except the first one are below the 
sums of $p^\downarrow_{+,n}$. The case $\bar{n} = 1$ is where the difference of partial sums is maximal. 
Therefore, the classicality criterion $p_+ \prec p_- $ is not satisfied in this case, which otherwise is an 
example of classical light according to most of other criteria, with the exceptions in Ref. \cite{nc}. This 
example also illustrates the lack of equivalence between clustering and Poissonian  behavior.

\begin{figure}
\begin{center}
\includegraphics[width=8cm]{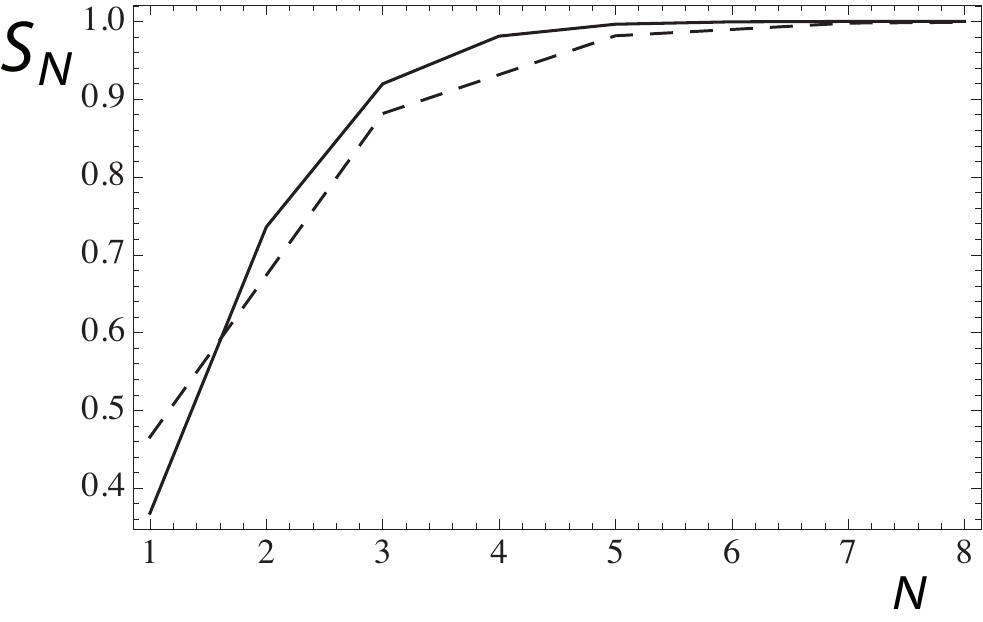}
\caption{Ordered partial sums for the number-sum distribution $p_+$ (solid line) and the number 
difference distribution $p_-$ (dashed line) for a coherent state with $\bar{n}=1$.}
\end{center}
\end{figure}

Furthermore, we have applied criteria (\ref{bc}) to the incomparable cases in Fig. 3 and Fig. 5 always 
obtaining majorization relations that agree with the comparison of their variances. We have obtained more 
interesting results examining the super-Poissonian squeezed state with $\bar{n}=6$ and  $\Delta^2 n 
= 2 \bar{n}$. Despite variance, this state displays an effective over-Poissonian behavior (less uncertainty 
than a coherent state of the same $\bar{n}$) for confidence intervals up to $\alpha \simeq 0.9$, as 
illustrated in Fig. 8. This agrees with the ordered partial sums for $p_\pm$ in Fig. 9 that show that this 
state presents clear anti-clustering (less uncertainty in the number sum than in the number difference) 
for all confidence intervals with $\alpha$ values up to $\alpha =0.9$.

\begin{figure}
\begin{center}
\includegraphics[width=8cm]{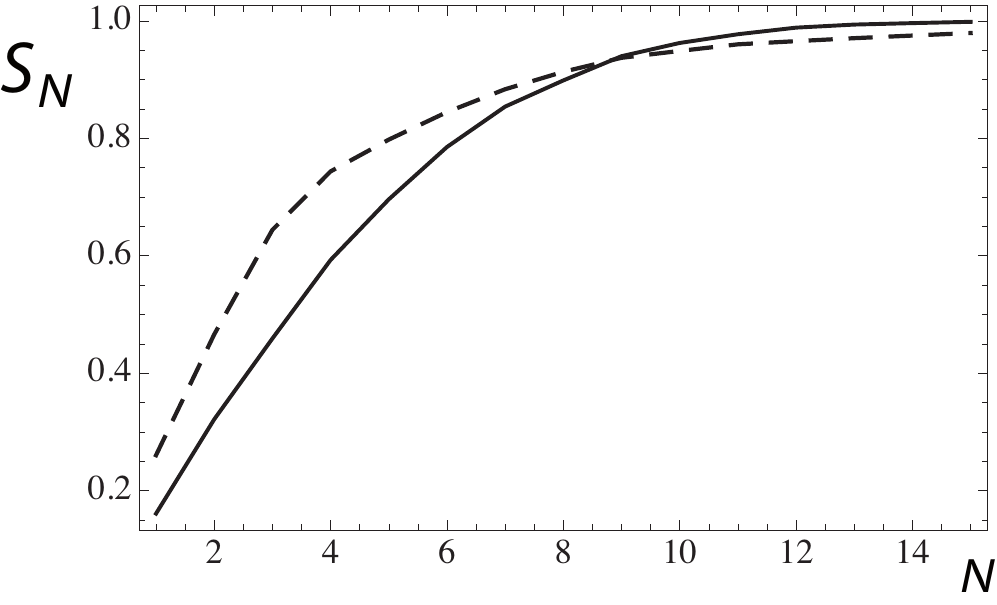}
\caption{Ordered partial sums for the photon-number distribution of coherent squeezed state with 
super-Poissonian variance $\Delta^2 n = 2 \, \bar{n}$ (dashed) and coherent light $\Delta^2 n =\bar{n}$ 
(solid) as functions of $N$ with $\bar{n}=6$. There is a clear effective over-Poissonian behavior 
(less uncertainty than a coherent  state of the same $\bar{n}$)  for confidence intervals up to $\alpha 
\simeq 0.9$. This contradicts the comparison of their variances.}
\end{center}
\end{figure}

\begin{figure}
\begin{center}
\includegraphics[width=8cm]{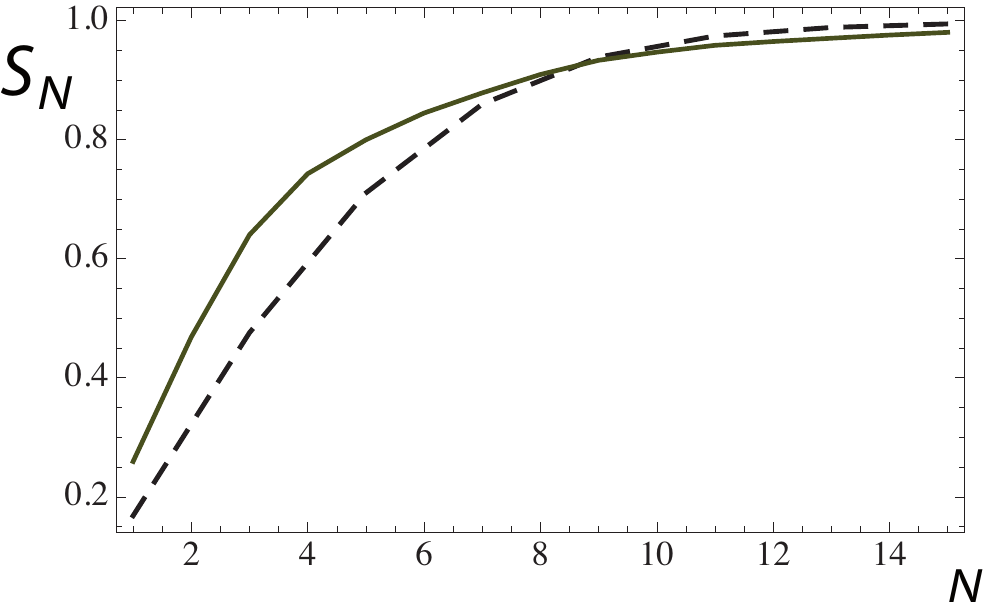}
\caption{Ordered partial sums for the number-sum distribution $p_+$ (solid line) and the number 
difference distribution $p_-$ (dashed line) for the coherent squeezed state with super-Poissonian 
variance  $\Delta^2 n = 2 \, \bar{n}$ and $\bar{n}=6$. There is a clear effective anti-clustering effect 
(less uncertainty in the number sum than in the number difference) for confidence intervals up to $\alpha 
\simeq 0.9$. This contradicts the conclusions derived from variance. }  
\end{center}
\end{figure}

In order to find further results contradicting the conclusions derived from variance let us consider the 
following mix of one-photon $| 1 \rangle$ and thermal $\rho_t$ states
\begin{equation}
\label{ptt}
\rho = \xi | 1 \rangle \langle 1 | + ( 1 - \xi ) \rho_t , \qquad 
\rho_t = \frac{1}{1+\bar{n}_t } \sum_{n=0}^\infty \left ( \frac{\bar{n}_t}{1+\bar{n}_t } \right )^n | n \rangle 
\langle n | ,
\end{equation}
where $\xi$ is a mixing parameter and $\bar{n}_t$ is the mean number of the thermal state $\rho_t$.
The complete state $\rho$ which has a mean number of photons of $\bar{n} = \xi + (1 - \xi ) \bar{n}_t $. 
For example for $\bar{n}= 2$ and  $\xi = 0.9$  the state $\rho$ is clearly super-Poissonian $\Delta^2 n 
= 11 \bar{n}$. However the whole statistics tell the full opposite. This is a consequence of the fact that
the number distribution is strongly concentrated around $n=1$, with $p_1 = 0.91$, so for most practical 
purposes this works as a single-photon state, that is, over-Poissonian and anti-clustered. This is clearly 
reflected in Fig. 10 when comparing with a coherent state of the same mean number. The states are 
incomparable, but when referring to confidence intervals enclosing up to 90 \% of the statistics $\alpha 
\leq  0.9$ the state behaves effectively as over-Poissonian (less fluctuations than the coherent state) in spite 
of $\Delta^2 n = 11 \bar{n}$. This is again the effect on variance of large photon numbers with negligible 
probabilities.

Regarding anti-clustering, the ordered partial sums of the distributions $p_\pm$ are represented in Fig. 11.
They show that the states are incomparable. But again,  for confidence intervals with $\alpha \leq 0.9$  the 
state behaves as anti-clustered (less uncertainty in the number sum than in the number difference), 
in contradiction with the predictions of variance. In particular the probability that only one detector registers 
light \textemdash this is the probability of $n_1 n_2 =0$ \textemdash is above $92 \%$.

\begin{figure}
\begin{center}
\includegraphics[width=8cm]{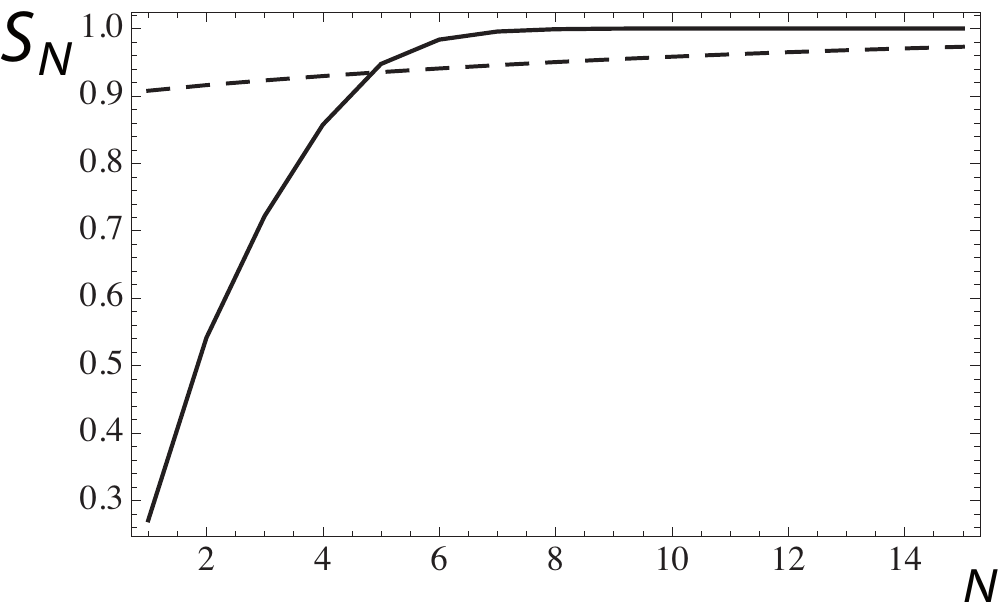}
\caption{Ordered partial sums for the photon times thermal state (\ref{ptt}) with  $\bar{n}= 2$ and  
$\xi = 0.9$ in dashed line and a coherent state of the same mean number in solid line.}
\end{center}
\end{figure}

\begin{figure}
\begin{center}
\includegraphics[width=8cm]{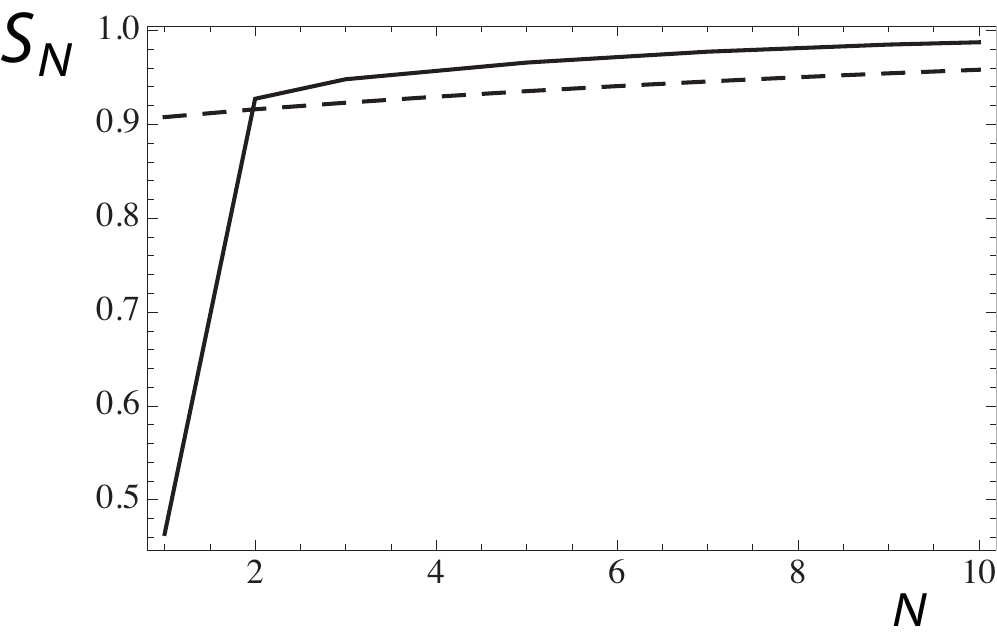}
\caption{Ordered partial sums for the number-sum distribution $p_+$ (solid line) and the number 
difference distribution $p_-$ for the photon times thermal state (\ref{ptt}) with  $\bar{n}= 2$ and  
$\xi = 0.9$.}  
\end{center}
\end{figure}

\section{Conclusions}

We have compared the photon statistics of the most relevant classical and nonclassical states from 
the perspective of majorization and confidence intervals. We have focused on the ideas of sub- and 
super-Poissonian statistics as well as the bunching and anti-bunching effects. 

As the most relevant results, we have found: (i) There are several interesting and common 
examples of incomparable states. (ii) Majorization may contradict variance. (iii) There is a 
majorization counterpart of bunching which is  logically independent of Poissonian statistics. 
Let us briefly comment on these results.

The information provided by majorization is valuable even when dealing with incomparable 
states. Previous works have found that different entropies lead to contradictory conclusions, 
where the minimum uncertainty states with one measure are the maximum uncertainty states 
of the other \cite{CD}. Majorization clearly explains this result showing that these 
contradictions arise because we are dealing with incomparable states \cite{CD2}.

Majorization can contradict variance because the former involves the complete statistics (that is, 
moments of all orders of the distribution), while variance only involves the first two moments.

In any case,  majorization will always provide a deeper knowledge of the situation than focusing 
exclusively on the variance. We do not propose to adopt majorization as a better way of comparing 
statics, but as another tool to understand quantum features. The use of one measure of the other 
will ultimately depend on the particular tasks addressed and the characteristic of the data available.

\ack

A. L. acknowledges support from Projects No. FIS2012-35583 of the Spanish Ministerio de Econom\'{\i}a 
y Competitividad and CAM research consortium QUITEMAD+ S2013/ICE-2801.

\end{document}